# A Função de Distribuição de Velocidades de Maxwell-Boltzmann em Detalhes

## (The Maxwell-Boltzmann Velocity Distribution Function in Detail)

Gilberto Jonas Damião, Clóves Gonçalves Rodrigues
Pontifícia Universidade Católica de Goiás, Goiânia, Brazil

**Resumo**

Apesar de sua importância, nas disciplinas introdutórias dos cursos de ciências exatas, a demonstração da lei de distribuição de velocidades de Maxwell-Boltzmann não é explicada, mostrando-se apenas a sua equação final. A fim de preencher esta deficiência, neste trabalho procuramos mostrar em detalhes, de forma bem didática a demonstração de tal lei. Para isto é inicialmente introduzida a teoria cinética dos gases. Apresenta-se também a boa concordância da lei de distribuição de velocidades de Maxwell-Boltzmann com resultados experimentais e o seu limite de aplicabilidade.

Palavras-chave: Cinética dos gases, distribuição de Maxwell-Boltzmann, velocidade média quadrática.

**Abstract**

Despite its importance, in the introductory disciplines of exact science courses, the demonstration of the Maxwell-Boltzmann velocity distribution law is not explained, only its final equation is shown. In order to fill this deficiency, in this work we try to show in detail, in a very didactic way, the demonstration of such a law. For this, the kinetic theory of gases is initially introduced. The good agreement of the Maxwell-Boltzmann velocity distribution law with experimental results and its applicability limit is also presented.

Keywords: Gas kinetics, Maxwell-Boltzmann distribution, mean square velocity.

## 1. Introdução

A temperatura de qualquer sistema físico é o resultado do movimento das moléculas e átomos que compõem o sistema. Essas pequenas partes da matéria possuem um intervalo de diferentes velocidades, e a velocidade de cada partícula varia constantemente devido a colisões umas com as outras. Tal distribuição relativa às velocidades, especifica a fração para cada intervalo de velocidades como função da temperatura do sistema. Esta distribuição leva os nomes de seus criadores: o físico e matemático britânico James Clerk Maxwell (1831-1879) e o físico austríaco Ludwig Eduard Boltzmann (1844-1906). No início da segunda metade do século XIX, por volta de 1859 Maxwell realizou estudos sobre como se distribuíam os módulos das velocidades das moléculas de um gás em equilíbrio térmico e no ano de 1860 demonstrou e divulgou que as velocidades das moléculas de um gás são distribuídas segundo a lei das distribuições dos erros, que foi formulada no ano de 1795 pelo matemático, físico e astrônomo alemão Johann Carl Friedrich Gauss (1777-1855). Nesta lei a energia cinética das moléculas é proporcional à temperatura absoluta $T$ do gás. Posteriormente, no ano de 1872 Boltzmann generalizou esta lei, sendo atualmente conhecida como "lei de Maxwell-Boltzmann".

A distribuição de velocidades de Maxwell-Boltzmann é em suma uma distribuição de probabilidade a qual pode ser aplicada em diversas áreas como física, engenharia, biologia, química, etc. O processo de evaporação da água em um lago, o brilho do sol, nêutrons térmicos em reatores nucleares, são todos exemplos de fenômenos em que a distribuição de velocidades de Maxwell-Boltzmann pode ser empegada.

Apesar de sua importância, os cursos na área de ciências exatas não tratam sobre a origem e a demonstração da lei de distribuição de velocidades de Maxwell-Boltzmann, apresentando apenas a sua equação final. A fim de preencher esta deficiência, neste trabalho procuramos mostrar em detalhes, de forma bem didática a demonstração de tal lei. Para isto introduzimos inicialmente na Seção 2 a teoria cinética dos gases antes de iniciar a demonstração da lei de Maxwell-Boltzmann, a qual é feita na Seção 3. Por fim, a Seção 4 se destina a questão da verificação experimental da lei de distribuição de velocidades de Maxwell-Boltzmann e o seu limite de validade.

## 2. A Teoria Cinética dos Gases

O movimento das partículas de um gás é muito complexo de ser analisado, pois envolve um número imenso de partículas com suas respectivas posições e momentos lineares. Normalmente esse número é da ordem do número de Avogadro[†], o que inviabiliza o cálculo direto de grandezas físicas individuais como a energia cinética e a velocidade.

Nesse caso, para se extrair informações físicas relevantes do modelo, utiliza-se uma abordagem estatística das grandezas individuais daqueles corpos microscópicos. Essas grandezas são chamadas de "variáveis ou quantidades microscópicas". Estas, por sua vez, implicam efeitos macroscópicos mensuráveis na forma de pressão, temperatura e volume – aspectos envolvidos na equação de estado de determinada substância (SCHIFINO, 2013). Essas grandezas são chamadas de "variáveis ou quantidades macroscópicas".

Conceitualmente, a pressão de um gás é a razão entre a força exercida por suas partículas nas faces internas das paredes de um recipiente e a área superficial interna do recipiente. A temperatura do gás é a medida da energia cinética média de seus átomos ou moléculas (SERWAY, 2014), e o seu volume é o volume do recipiente que o contém.

A energia cinética média $\langle E \rangle$ leva em consideração o valor médio do quadrado da velocidade das partículas do gás, $\langle v^2 \rangle$, a partir da seguinte relação

$$\langle E \rangle = \frac{m \langle v^2 \rangle}{2},$$ (1)

onde $m$ é a massa de seus átomos ou moléculas (SERWAY, 2014).

Busca-se, então, um modelo físico-matemático que relacione os parâmetros macroscópicos temperatura ($T$), pressão ($P$) e volume do gás ($V$) com a sua energia cinética média e velocidade média quadrática das partículas.

Consideremos um gás a uma temperatura constante contido em um recipiente cúbico. A pressão que suas partículas exercem nas paredes internas depende da transferência de momento na forma de colisões elásticas entre as partículas que constituem o gás e as faces internas das paredes do recipiente que o contém (SERWAY, 2014). Neste modelo, o gás é composto por átomos ou moléculas puntiformes e as forças oriundas da interação entre os corpúsculos, bem como suas colisões, são desprezíveis.

Uma partícula isolada tem massa $m$ e velocidade $\vec{v}$ com as direções e módulos das velocidades livres. As partículas estão submetidas às leis de movimento de Newton com um movimento total isotrópico. A Figura 1 ilustra o choque elástico de uma partícula (átomo/molécula) com uma das paredes do recipiente.

---

[†] Quantidade de partículas por mol de determinada substância, que tem o valor aproximado de $6{,}022 \times 10^{23}$ unidades, correspondente a mais de seiscentos bilhões de trilhões de unidades, ou $602{,}2 \times 10^9 \times 10^{12}$ unidades.

**Figura 1:** *Molécula contida em um recipiente cúbico.*

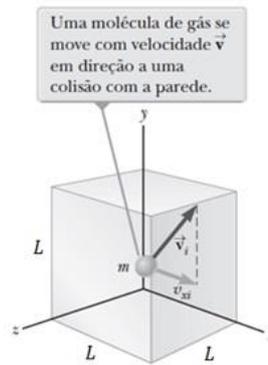

A transferência de momento da parede para a partícula se dá, matematicamente, na direção $x$, da seguinte forma

$$p_{x(\text{inicial})} = mv_{x(\text{inicial})}\,,\tag{2}$$

e

$$p_{x(\text{final})} = mv_{x(\text{final})}\,,\tag{3}$$

e considerando a colisão perfeitamente elástica temos

$$v_{x(\text{final})} = -v_{x(\text{inicial})}.\tag{4}$$

Assim, a variação do momento linear[*] da partícula imediatamente após a colisão com a parede é

$$\Delta p = p_{x(\text{final})} - p_{x(\text{inicial})} = mv_{x(\text{final})} - mv_{x(\text{inicial})}\,,$$

Usando a Eq. (4), essa última expressão fica

$$\Delta p = -mv_{x(\text{inicial})} - mv_{x(\text{inicial})} = -2mv_{x(\text{inicial})} = -2mv_x\,,$$

onde foi definindo $v_x = v_{x(\text{inicial})}$. Portanto, o momento transferido da molécula do gás para a parede interna é

$$\Delta p = 2mv_x\,.\tag{5}$$

O tempo necessário para que a partícula atinja a parede frontal e retorne, considerando um cubo de aresta $L$ é

$$v_x = \frac{2L}{\Delta t} \Rightarrow \Delta t = \frac{2L}{v_x}\,.\tag{6}$$

A variação do momento linear $p$ em relação à variação temporal consiste na grandeza física força (2ª Lei de Newton), isto é

$$\frac{\Delta p}{\Delta t} = F \Rightarrow \Delta p = F\Delta t,$$

e usando a Eq. (5) esta última expressão fica

$$F\Delta t = 2mv_x \Rightarrow F = \frac{2mv_x}{\Delta t}\,.\tag{7}$$

---

[*] utilizaremos $p$ (minúsculo) para representar o momento e $P$ (maiúsculo) para representar a pressão.

Substituindo a Eq. (6) na Eq. (7) temos

$$F = \frac{2mv_x}{2L/v_x} \Rightarrow F = \frac{mv_x^2}{L}.$$ (8)

Esta é a expressão da força que uma partícula produz na parede do recipiente. Para encontrar a pressão $P$ exercida nesta face da parede, deve-se, primeiro, encontrar a força que todas as $N$ partículas aplicam nela e relacioná-la com a sua área $L^2$

$$F_{total} = \sum_{i=1}^{N} \frac{mv_{xi}^2}{L} \Rightarrow P = \frac{F_{total}}{L^2} = \frac{1}{L^2} \sum_{i=1}^{N} \frac{mv_{xi}^2}{L} = \frac{m}{L^3} \sum_{i=1}^{N} v_{xi}^2.$$

A média do quadrado das velocidades na direção $x$ pode ser encontrada com a multiplicação nesta última expressão pela fração $N/N$:

$$P = \left(\frac{N}{N}\right)\frac{m}{L^3}\left(\sum_{i=1}^{N} v_{xi}^2\right) = \frac{mN}{L^3}\left(\sum_{i=1}^{N} \frac{v_{xi}^2}{N}\right) = \frac{mN}{L^3}\langle v_x^2 \rangle$$

$$P = \frac{mN}{V}\langle v_x^2 \rangle,$$ (9)

onde $L^3$ é o volume $V$ do cubo.

As velocidades médias independem da direção adotada, pois há um número muito grande de partículas se movendo aleatoriamente de forma muito rápida (YOUNG, 2008). Dessa forma, os módulos das velocidades médias são iguais em qualquer direção. Assim

$$\langle v^2 \rangle = \langle v_x^2 \rangle + \langle v_y^2 \rangle + \langle v_z^2 \rangle,$$ (10)

com

$$\langle v_x^2 \rangle = \langle v_y^2 \rangle = \langle v_z^2 \rangle,$$ (11)

tem-se

$$\langle v^2 \rangle = 3\langle v_x^2 \rangle \Rightarrow \langle v_x^2 \rangle = \frac{1}{3}\langle v^2 \rangle.$$ (12)

Substituindo (12) em (9) temos

$$P = \frac{mN}{V}\frac{\langle v^2 \rangle}{3} \Rightarrow PV = \frac{mN}{3}\langle v^2 \rangle.$$ (13)

A equação dos gases ideais (HALLIDAY, 2012) dada por

$$PV = nRT,$$ (14)

onde $R$ é a constante dos gases ideais e $n$ o número de mols, pode ser relacionada com a Eq. (13) da seguinte forma

$$nRT = \frac{mN}{3}\langle v^2 \rangle.$$

Sendo $N = nN_A$, onde $N_A$ é o número de Avogadro, a expressão anterior assume a forma

$$nRT = \frac{mnN_A}{3}\langle v^2 \rangle \Rightarrow RT = \frac{mN_A}{3}\langle v^2 \rangle.$$

Usando que a massa molar $M$ do gás é $M = mN_A$, a expressão anterior fica

$$RT = \frac{M}{3}\langle v^2 \rangle. \tag{15}$$

Isolando $\langle v^2 \rangle$, temos

$$\langle v^2 \rangle = \frac{3RT}{M},$$

e extraindo a sua raiz quadrada, temos a raiz quadrada do valor quadrático médio da velocidade das partículas, denotada por $v_{\mathrm{rms}}$ (uma acrossemia do inglês: *root mean square* = rms) (SCHIFINO, 2013) dada, então, por:

$$v_{\mathrm{rms}} = \sqrt{\frac{3RT}{M}} \tag{16}$$

ou

$$v_{\mathrm{rms}} = \sqrt{\frac{3kT}{m}}, \tag{17}$$

onde $R/N_A = k = 1,3806 \times 10^{-23}$ J/K é a constante de Boltzmann (HALLIDAY, 2012). Pode-se afirmar, também, que a velocidade quadrática média de cada uma das componentes $v_x$, $v_y$ e $v_z$ é, a partir da Eq. (15)

$$RT = \frac{M}{3}\langle v^2 \rangle = \frac{M}{3}3\langle v_x{}^2 \rangle,$$

$$\langle v_x{}^2 \rangle = \frac{RT}{M}, \tag{18}$$

ou

$$\langle v_x{}^2 \rangle = \frac{kT}{m}. \tag{19}$$

Logo

$$v_{x(\mathrm{rms})} = \sqrt{\frac{RT}{M}}, \tag{20}$$

ou

$$v_{x(\mathrm{rms})} = \sqrt{\frac{kT}{m}}. \tag{21}$$

## 3. A Função de Distribuição de Velocidades de Maxwell-Boltzmann

Os resultados obtidos na Seção 2 são muito significativos, pois, a partir de uma grandeza física macroscópica como a temperatura $T$ é possível determinar a velocidade $v_{\mathrm{rms}}$ com a qual se movem as partículas do gás. Entretanto, a dedução exposta leva a outra hipótese.

Se existe $v_{rms}$, parâmetro estatístico da velocidade, pode existir uma distribuição de prováveis velocidades, isto é, uma função densidade de probabilidade que descreva os intervalos de velocidades a serem assumidos pelos corpúsculos deste gás.

O físico britânico James Clerk Maxwell apresentou, em 1859, um trabalho no qual expôs uma distribuição de velocidades de um gás em equilíbrio térmico (NUSSENZVEIG, 2002) e, em 1876, Ludwig Boltzmann chegou ao mesmo resultado por um modelo diferente.

A seguir apresenta-se uma possível dedução desta função de distribuição de velocidades.

Consideremos um espaço tridimensional de velocidades moleculares, como ilustrado na Figura 2. O objetivo é encontrar a probabilidade das partículas de um gás, que possuem velocidades diversas, apresentarem componentes de velocidade entre $v_x$ e $v_x + dv_x$; $v_y$ e $v_y + dv_y$; e $v_z$ e $v_z + dv_z$ (NUSSENZVEIG, 2002).

**Figura 2:** *Espaço de velocidades.*

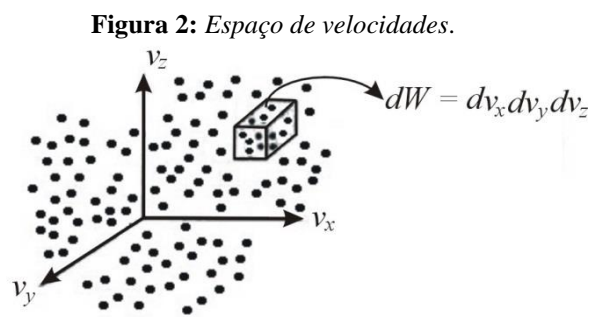

Seja $f(v_x, v_y, v_z)dv_x dv_y dv_z$ uma fração do total de partículas de um gás com componentes $v_x$, $v_y$ e $v_z$ compreendidas na faixa de velocidades exposta no parágrafo anterior. Deseja-se descobrir qual é esta função de distribuição de velocidades. Como a velocidade quadrática média depende da temperatura do gás, supõe-se que esta distribuição também dependa. Considere-se o gás em equilíbrio térmico, com a temperatura $T$ constante (SCHIFINO, 2013).

Pela definição de probabilidade de variáveis aleatórias contínuas (DEVORE, 2006), e considerando que o número de moléculas é suficientemente grande para que se adotem como contínuos os valores de velocidades contidas num intervalo cujos limites tendam a $-\infty$ e $+\infty$ em cada dimensão do espaço de velocidades, tem-se

$$\int_{-\infty}^{+\infty} \int_{-\infty}^{+\infty} \int_{-\infty}^{+\infty} f(v_x, v_y, v_z)dv_x dv_y dv_z = 1, \tag{22}$$

ou

$$\int_{-\infty}^{+\infty} \int_{-\infty}^{+\infty} \int_{-\infty}^{+\infty} f(v_x, v_y, v_z)dW = 1, \tag{23}$$

sendo $dW$ o elemento de volume do espaço de velocidades igual a $dv_x dv_y dv_z$.

Contudo, duas considerações devem ser feitas, similares às utilizadas na Seção 2 para a dedução da velocidade $v_{rms}$. Em primeiro lugar, as componentes são independentes, ou seja, a velocidade $v_x$, por exemplo, não está vinculada aos valores de $v_y$ e $v_z$. Isso é válido para cada componente em relação às outras duas. Em segundo lugar, não há direção espacial preferencial para onde as moléculas caminham, ou seja, o espaço é considerado isotrópico (BAGNATO, 2016).

Matematicamente, estes princípios se expressam com a adoção da mesma forma funcional para cada componente (isotropia espacial) e com a separação da função principal em funções autônomas em relação às outras (independência entre as componentes), isto é

$$\int_{-\infty}^{+\infty} f(v_x) dv_x = 1, \tag{24}$$

$$\int_{-\infty}^{+\infty} f(v_y) dv_y = 1, \tag{25}$$

$$\int_{-\infty}^{+\infty} f(v_z) dv_z = 1. \tag{26}$$

Como essas componentes são independentes e suas funções são interpretadas probabilisticamente, a função $f(v_x, v_y, v_z) dv_x dv_y dv_z$ é igual ao produto das funções componentes, ou seja

$$f(v_x, v_y, v_z) dv_x dv_y dv_z = f(v_x) dv_x f(v_y) dv_y f(v_z) dv_z,$$

e integrando em todo o espaço

$$\int_{-\infty}^{+\infty} \int_{-\infty}^{+\infty} \int_{-\infty}^{+\infty} f(v_x, v_y, v_z) dv_x dv_y dv_z = \int_{-\infty}^{+\infty} f(v_x) dv_x \int_{-\infty}^{+\infty} f(v_y) dv_y \int_{-\infty}^{+\infty} f(v_z) dv_z. \tag{27}$$

A isotropia espacial implica a não dependência da direção que se adota, pois a probabilidade de se encontrar determinado número de partículas em uma fração do espaço de velocidades deve ser a mesma que a de outra região equidistante da origem daquele espaço. Portanto, faz-se necessário considerar não mais as componentes da velocidade, mas o módulo desta na função distribuição de velocidades (BAGNATO, 2016), isto é

$$f(v) dW = f(v_x) f(v_y) f(v_z) dW, \tag{28}$$

onde $v = \sqrt{v_x^2 + v_y^2 + v_z^2}$. Assim

$$f(v) = f(v_x) f(v_y) f(v_z). \tag{29}$$

Tomando a derivada de $f(v)$ em relação a $v_x$, tem-se

$$\frac{df(v)}{dv_x} = \frac{df(v_x)}{dv_x} f(v_y) f(v_z)$$

$$\left(\frac{\partial f(v)}{\partial v}\right)\left(\frac{\partial v}{\partial v_x}\right) = \frac{df(v_x)}{dv_x}f(v_y)f(v_z) \,. \tag{30}$$

Como $v = (v_x^2 + v_y^2 + v_z^2)^{1/2}$, tem-se:

$$\frac{\partial v}{\partial v_x} = \frac{1}{2}(v_x^2 + v_y^2 + v_z^2)^{-1/2} \cdot 2v_x \;=\; \frac{v_x}{\sqrt{v_x^2 + v_y^2 + v_z^2}} = \frac{v_x}{v} \tag{31}$$

Além disso

$$\frac{\partial f(v)}{\partial v} = f'(v) \tag{32}$$

e

$$\frac{df(v_x)}{dv_x} = f'(v_x) \,. \tag{33}$$

Inserindo (31), (32) e (33) na Eq. (30), temos

$$\frac{f'(v)}{v}v_x = f'(v_x)f(v_y)f(v_z) \,. \tag{34}$$

Nota-se que o quociente entre (34) e (29) promove o cancelamento do termo $f(v_y)f(v_z)$ da seguinte forma

$$\frac{f'(v)v_x}{f(v)v} = \frac{f'(v_x)}{f(v_x)}$$

$$\frac{f'(v)}{f(v)v} = \frac{f'(v_x)}{f(v_x)v_x} \,. \tag{35}$$

Neste ponto, cabe uma análise da Eq. (35). Enquanto o lado direito da igualdade é função que depende de uma das componentes da velocidade, verifica-se que o lado esquerdo vincula-se somente ao módulo da velocidade $v$. Por sua vez, o módulo da velocidade subordina-se aos valores de $v_x$, $v_y$ e $v_z$. Então, para um dado valor de $v_x$, há infinitas combinações de $v_y$ e $v_z$ que tornam a igualdade acima válida. Dessa forma, a relação acima só faz sentido matemático se for igual a um valor constante (BAGNATO, 2016), isto é

$$\frac{f'(v)}{f(v)v} = \frac{f'(v_x)}{f(v_x)v_x} = \frac{f'(v_y)}{f(v_y)v_y} = \frac{f'(v_z)}{f(v_z)v_z} = \text{constante} \,. \tag{36}$$

Faremos esta constante igual a $-2\alpha$ em virtude da maior facilidade em se trabalhar com o expoente negativo após a integração de uma das equações acima. Assim

$$\frac{f'(v)}{f(v)v} = -2\alpha \tag{37}$$

e

$$\frac{f'(v_x)}{f(v_x)v_x} = -2\alpha \,. \tag{38}$$

Substituindo (33) em (38)

$$\frac{df(v_x)}{dv_x f(v_x) v_x} = -2\alpha$$

$$\frac{df(v_x)}{f(v_x)} = -2\alpha v_x dv_x \,. \tag{39}$$

Integrando a Eq. (39)

$$\int \frac{df(v_x)}{f(v_x)} = \int -2\alpha v_x dv_x \Rightarrow \ln\big(f(v_x)\big) = -\alpha v_x^2 + C \Rightarrow f(v_x) = e^{-\alpha.v_x^2 + C},$$

$$f(v_x) = Ae^{-\alpha.v_x^2} \,, \tag{40}$$

onde $A = e^C$. A função da Eq. (40) é uma curva gaussiana semelhante à distribuição normal (DEVORE, 2006). De posse desta relação, pode-se encontrar $f(v)$, considerando que as funções das outras componentes do espaço de velocidades são

$$f(v_y) = Ae^{-\alpha.v_y^2}, \tag{41}$$

e

$$f(v_z) = Ae^{-\alpha.v_z^2}. \tag{42}$$

Substituindo (40), (41) e (42) na Eq. (29) tem-se

$$f(v) = Ae^{-\alpha.v_x^2}Ae^{-\alpha.v_y^2}Ae^{-\alpha.v_z^2} = A^3 e^{-\alpha.(v_x^2 + v_y^2 + v_z^2)}$$

$$f(v) = A^3 e^{-\alpha.v^2}. \tag{43}$$

A partir da Eq. (24) e da Eq. (19) é possível determinar as constantes $A$ e $\alpha$. Assim, inserindo (40) em (24) tem-se

$$I_1 = \int\limits_{-\infty}^{+\infty} Ae^{-\alpha.v_x^2}\, dv_x = 1, \tag{44}$$

e com procedimento análogo em relação à componente $y$

$$I_2 = \int\limits_{-\infty}^{+\infty} Ae^{-\alpha.v_y^2}\, dv_y = 1 \,. \tag{45}$$

O produto entre a Eq. (44) e a Eq. (45) resulta em

$$I_1 \cdot I_2 = \int\limits_{-\infty}^{+\infty} \int\limits_{-\infty}^{+\infty} A^2 e^{-\alpha(v_x^2 + v_y^2)}\, dv_y dv_x = 1. \tag{46}$$

Observa-se pela simetria das integrais, que $I_1 = I_2 = I$. Tomando um plano cartesiano de abscissa $v_x$ e ordenada $v_y$, podem-se utilizar as seguintes relações para converter tais coordenadas em polares (RODRIGUES, 2017)

$$v_x^2 + v_y^2 = r^2, \tag{47a}$$

$$v_x = r\cos\theta, \tag{47b}$$

$$v_y = r\sin\theta, \tag{47c}$$

$$dv_y dv_x = \det J = \begin{vmatrix} \cos(\theta)dr & -r\sin(\theta)d\theta \\ \sin(\theta)dr & r\cos(\theta)d\theta \end{vmatrix} = r\,dr\,d\theta\,. \tag{47d}$$

Os limites de integração $-\infty < v_x < +\infty$ e $-\infty < v_y < +\infty$ das coordenadas cartesianas são equivalentes aos limites $0 \le r < \infty$ e $0 \le \theta \le 2\pi$ das coordenadas polares. Assim, com as definições apresentadas nas Eqs. (47a-d), a Eq. (46) assume a forma

$$I^2 = I_1 . I_2 = \int_0^{2\pi} \int_0^{+\infty} A^2 e^{-\alpha r^2}\, r\,dr\,d\theta = 1,$$

$$I^2 = A^2 \int_0^{2\pi} d\theta \int_0^{+\infty} e^{-\alpha r^2} r\,dr = 1,$$

$$I^2 = A^2 \cdot 2\pi \cdot \frac{1}{2\alpha} = A^2 \frac{\pi}{\alpha} = 1$$

$$I = A\sqrt{\frac{\pi}{\alpha}} = 1 \Rightarrow A\sqrt{\frac{\pi}{\alpha}} = 1$$

$$A = \sqrt{\frac{\alpha}{\pi}}\,. \tag{48}$$

Para determinar a constante $\alpha$, utiliza-se a expressão de $\langle v_x^2 \rangle$ obtida na Seção 2. Conceitualmente, a média quadrática de uma distribuição contínua, a exemplo de $\langle v_x^2 \rangle$, se dá pela integral

$$\langle v_x^2 \rangle = \int_{-\infty}^{+\infty} v_x^2 f(v_x) dv_x,$$

e usando a expressão de $\langle v_x^2 \rangle$ dada em (19), a equação anterior fica

$$\frac{kT}{m} = \int_{-\infty}^{+\infty} v_x^2 f(v_x) dv_x,$$

$$\int\limits_{-\infty}^{+\infty} v_x^2 \, e^{-\alpha v_x^2} dv_x = \frac{kT}{Am}.$$

Substituindo o valor de $A$ dado pela Eq. (48) em e o valor tabelado da integral (GRADSHTEYN, 2007) esta última equação fica

$$\frac{1}{2}\sqrt{\frac{\pi}{\alpha^3}} = \frac{kT}{\sqrt{\alpha/\pi m}},$$

e isolando $\alpha$

$$\alpha = \frac{m}{2kT}. \tag{49}$$

Introduzindo (49) em (48) tem-se

$$A = \sqrt{\frac{m}{2\pi kT}}. \tag{50}$$

Introduzindo as Eqs. (49) e (50) em (43), obtém-se

$$f(v) = \left(\sqrt{\frac{m}{2\pi kT}}\right)^3 e^{-\frac{mv^2}{2kT}}. \tag{51}$$

O produto de $f(v)$ pelo elemento de volume $dW$ possibilita generalizar tal modelo. Com isso, a função torna-se dependente apenas do módulo da velocidade (BAGNATO, 2016):

$$f(v)dW = f(v)dv_x dv_y dv_z = f(v)v^2 d\Omega dv \tag{52}$$

Note que

$$\int\limits_{0}^{4\pi} f(v)v^2 d\Omega dv = 4\pi v^2 f(v)dv. \tag{53}$$

Tem-se, portanto, a função de distribuição de velocidades preconizada por Maxwell e por Boltzmann, designada por $f_{MB}$

$$f_{MB}(v) = 4\pi \left(\sqrt{\frac{m}{2\pi kT}}\right)^3 v^2 e^{\frac{-mv^2}{2kT}}. \tag{54}$$

O gráfico desta distribuição assemelha-se a uma gaussiana com início na origem, crescimento quadrático e decrescimento exponencial (BAGNATO, 2016), como ilustrado na Figura 3. Nota-se que o efeito do aumento da temperatura torna a curva de probabilidade mais achatada e longa.

**Figura 3:** *Função de distribuição de velocidades de Maxwell-Boltzmann para dois valores de temperatura. O valor adotado para a massa $m$ foi a massa do próton.*

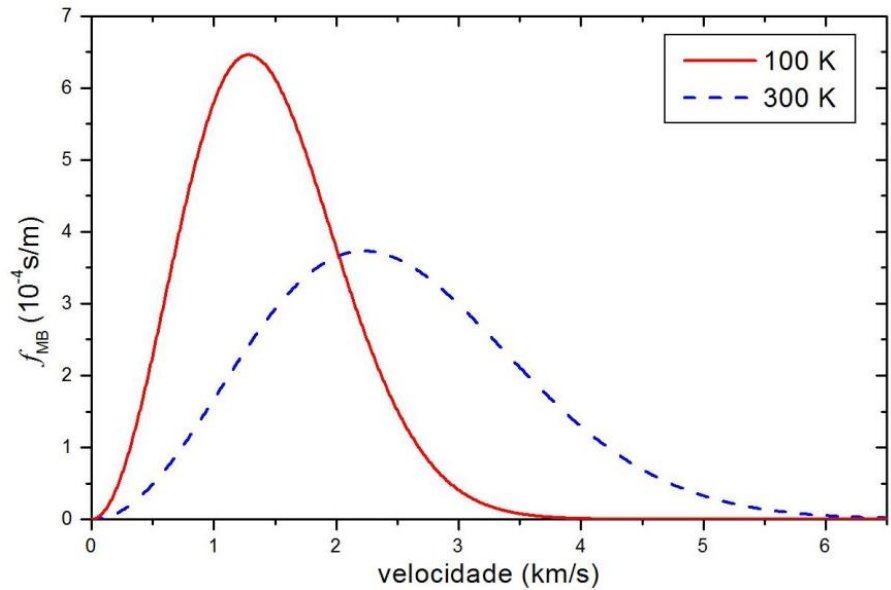

A interpretação, neste ponto, é de qual fração de partículas do gás possuem velocidade de módulo entre $v$ e $v + dv$, não importando a direção. Ela representa, também, a probabilidade de uma partícula apresentar velocidade entre $v$ e $v + dv$ (SERWAY, 2014). Portanto, não se pensa mais num cubo de dimensões $dv_x$, $dv_y$ e $dv_z$ que a contém, mas em uma casca esférica de espessura $dv$ no espaço de velocidades (NUSSENZVEIG, 2002; BAGNATO, 2016). A Figura 4 ilustra tal situação:

**Figura 4:** *Representação do espaço de velocidades com distribuição de magnitudes.*

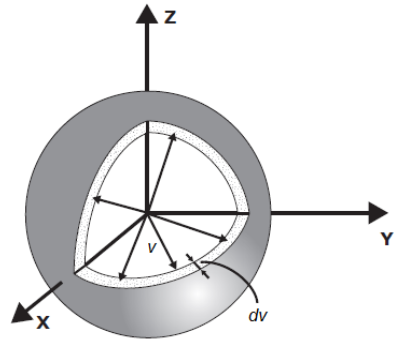

A velocidade mais provável, $v_p$, é o valor que maximiza a função $f_{MB}(v)$. Graficamente, seria o valor de velocidade responsável pelo pico da função de distribuição de velocidades. Portanto, tomando-se a derivada da distribuição de Maxwell-Boltzmann e igualando-a a zero, tem-se:

$$\frac{\partial(f_{MB}(v))}{\partial v} = \frac{\partial\left(4\pi\left(\sqrt{\dfrac{m}{2\pi kT}}\right)^3 v^2 e^{\frac{-mv^2}{2kT}}\right)}{\partial v} = 0,$$

$$2\sqrt{\frac{2}{\pi}}\left(\sqrt{\frac{m}{kT}}\right)^3 v_p e^{\frac{-m(v_p)^2}{2kT}} - \sqrt{\frac{2}{\pi}}\left(\sqrt{\frac{m}{kT}}\right)^3 \frac{m(v_p)^3}{kT} e^{\frac{-m(v_p)^2}{2kT}} = 0,$$

$$2\sqrt{\frac{2}{\pi}}\left(\sqrt{\frac{m}{kT}}\right)^3 v_p e^{\frac{-m(v_p)^2}{2kT}} = \sqrt{\frac{2}{\pi}}\left(\sqrt{\frac{m}{kT}}\right)^3 \frac{m(v_p)^3}{kT} e^{\frac{-m(v_p)^2}{2kT}},$$

e após algumas simplificações

$$2 = \frac{m(v_p)^2}{kT},$$

e finalmente

$$v_p = \sqrt{\frac{2kT}{m}} \ . \tag{55}$$

A velocidade média, $\langle v \rangle$, isto é, a esperança da velocidade desta distribuição é dada pela integral

$$\langle v \rangle = \int\limits_{0}^{+\infty} v f_{MB}(v) dv = \int\limits_{0}^{+\infty} v 4\pi \left(\sqrt{\frac{m}{2\pi kT}}\right)^3 v^2 e^{\frac{-mv^2}{2kT}} dv,$$

$$\langle v \rangle = 4\pi \left(\sqrt{\frac{m}{2\pi kT}}\right)^3 \int\limits_{0}^{+\infty} v^3 e^{\frac{-mv^2}{2kT}} dv,$$

e utilizando uma tabela de integrais (GRADSHTEYN, 2007), tem-se

$$\langle v \rangle = 4\pi \left(\sqrt{\frac{m}{2\pi kT}}\right)^3 \frac{4(kT)^2}{2m^2},$$

$$\langle v \rangle = \sqrt{\frac{8kT}{\pi m}} \ . \tag{56}$$

A velocidade $v_{\text{rms}}$, determinada na Seção 2, pode ser calculada a partir da distribuição de velocidades de Maxwell-Boltzmann da seguinte forma

$$\langle v^2 \rangle = \int\limits_{0}^{+\infty} v^2 f_{MB}(v) dv = \int\limits_{0}^{+\infty} v^2 4\pi \left(\sqrt{\frac{m}{2\pi kT}}\right)^3 v^2 e^{\frac{-mv^2}{2kT}} dv,$$

$$\langle v^2 \rangle = 4\pi \left(\sqrt{\frac{m}{2\pi kT}}\right)^3 \int\limits_{0}^{+\infty} v^4 e^{\frac{-mv^2}{2kT}} dv$$

e usando o valor tabelado da integral (GRADSHTEYN, 2007), tem-se

$$\langle v^2 \rangle = 4\pi \left(\sqrt{\frac{m}{2\pi kT}}\right)^3 \frac{3\sqrt{\pi(2kT)^5}}{8\sqrt{m^5}} = \frac{3kT}{m}$$

$$v_{\text{rms}} = \sqrt{\langle v^2 \rangle} = \sqrt{\frac{3kT}{m}} \ . \tag{57}$$

Observa-se que a Eq. (57) é igual à Eq. (17), o que evidencia a validade desta distribuição. Resumindo, foram obtidos:

- velocidade mais provável, $v_p$:

$$v_p = \sqrt{\frac{2kT}{m}}$$

- velocidade média, $\langle v \rangle$:

$$\langle v \rangle = \sqrt{\frac{8kT}{\pi m}}$$

- velocidade $v_{\text{rms}}$:

$$v_{\text{rms}} = \sqrt{\langle v^2 \rangle} = \sqrt{\frac{3kT}{m}}$$

A Figura 5 mostra a distribuição de velocidades de Maxwell-Boltzmann para uma temperatura de 300 K, destacando-se as velocidades: $v_p$, $\langle v \rangle$ e $v_{\text{rms}}$. Note que $v_p < \langle v \rangle < v_{\text{rms}}$.

**Figura 5:** Distribuição de velocidades de Maxwell-Boltzmann para uma temperatura de 300 K, destacando-se as velocidades: $v_p$, $\langle v \rangle$ e $v_{\text{rms}}$.

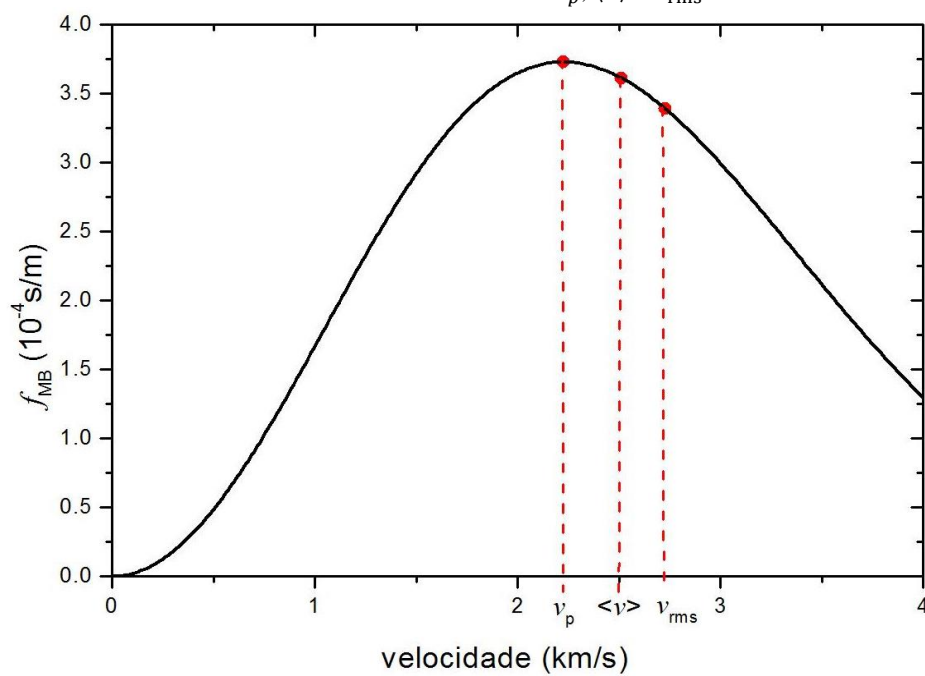

## 4. Comentários Finais

Uma maneira de verificar experimentalmente a validade da lei de distribuição de velocidades de Maxwell-Boltzmann consiste em analisar como variam alguns processos moleculares, como a velocidade das reações químicas quando a temperatura varia. Os valores experimentais obtidos nestes experimentos estão em excelente acordo com a teoria de Maxwell-Boltzmann.

Uma verificação mais direta da lei de distribuição de velocidades de Maxwell-Boltzmann consiste na contagem do número de moléculas em cada intervalo de velocidades ou de energias. Isto pode ser feito experimentalmente com um método que usa um seletor mecânico de velocidades composto por discos e fendas que giram com uma velocidade angular determinada que selecione as velocidades desejadas, por exemplo, em um tanque com um orifício contendo um gás a uma temperatura $T$. Os resultados experimentais novamente confirmam as predições de Maxwell-Boltzmann.

Nêutrons produzidos em processos de fissão de um reator nuclear são moderados por meio de um material, como água ou grafite, até atingirem o equilíbrio térmico à temperatura do moderador. Os nêutrons em equilíbrio térmico comportam-se como um gás ideal e a sua distribuição de energia concorda com a lei de distribuição de velocidades de Maxwell-Boltzmann, isto é, os nêutrons térmicos seguem a estatística de Maxwell-Boltzmann. Este fato é essencial na concepção de reatores nucleares (ALONSO, 1992).

Uma questão central é o fato da distribuição de Maxwell-Boltzmann admitir uma probabilidade não nula de se encontrar partículas com velocidades maiores que a velocidade da luz no vácuo $c$. No entanto, sabe-se pela teoria da relatividade que somente velocidades menores que $c$ possuem sentido físico. Portanto, analisaremos em um futuro artigo uma função de distribuição de velocidades capaz de lidar com este problema: a função de distribuição proposta por Jüttner (DUNKEL, 2007; HAKIM, 2011).